\documentclass[a4paper]{article}

\usepackage{INTERSPEECH2022}
\usepackage{multirow}
\usepackage{url}
\usepackage{color}
\title{The DKU-MSXF Speaker Verification System for the VoxCeleb Speaker Recognition Challenge 2023}
\name{Ze Li$^{1*}$,Yuke Lin$^{1*}$,Xiaoyi Qin$^{1*}$,  Ning Jiang $^{2}$, Guoqing Zhao$^{2}$, Ming Li$^{1}$\thanks{*These authors contributed equally to this work.}}
\address{
$^1$Data Science Research Center, Duke Kunshan University, Kunshan, China \\
$^2$Mashang Consumer Finance Co., Ltd, China
}
\email{ming.li369@dukekunshan.edu.cn}

\begin{document}

\maketitle

\begin{abstract}

This paper is the system description of the DKU-MSXF System for the track1, track2 and track3 of the VoxCeleb Speaker Recognition Challenge 2023 (VoxSRC-23). For Track 1, we utilize a network structure based on ResNet for training. By constructing a cross-age QMF training set, we achieve a substantial improvement in system performance. For Track 2, we inherite the pre-trained model from Track 1 and conducte mixed training by incorporating the VoxBlink-clean dataset. In comparison to Track 1, the models incorporating VoxBlink-clean data exhibit a performance improvement by more than 10\% relatively. For Track3, the semi-supervised domain adaptation task, a novel pseudo-labeling method based on triple thresholds and sub-center purification is adopted to make domain adaptation. The final submission achieves mDCF of 0.1243 in task1, mDCF of 0.1165 in Track 2 and EER of 4.952\% in Track 3.
\end{abstract}


\section{Track 1\&2:Fully supervised speaker verification}
\subsection{Data Usage}
For Track 1, we only adopt the VoxCeleb2 dev set \cite{vox2} with 1,092,009 utterances from 5994 speakers. For Track 2, we additionally incorporated data from VoxBlink-clean\cite{voxblink} into the training set, which contains 1,028,106 utterances from 18,381 speakers. The VoxBlink dataset is constructed by an automatic pipeline from short videos uploaded by users in Youtube. This dataset is one of the largest speaker recognition datasets in terms of the number of speakers. It also highlights cross-channel, cross-time, and cross-lingual characteristics, which will be released to public soon. Table \ref{tab:data_usage} illustrates the number of speakers and utterances of these two datasets.

\begin{table}[htbp]\centering
    \caption{Data usage for track1\&2. VoxBlink-clean is the purified version of the VoxBlink.}
     \label{tab:data_usage}
    \begin{tabular}{lccccccc}
    \toprule
        ID & Dataset & Spk. Num & Utt. Num\\
    \midrule 
        A & Vox2Dev & 5,994 & 1,092,009 \\
        B & VoxBlink-clean & 18,381 & 1,028,106 \\
    \bottomrule
    \end{tabular}
    \end{table}
    
\begin{table}[htbp]\centering \scriptsize
    \caption{The performance of a signal baseline system (SimAM-ResNet100).}
     \label{tab:baseline_system}
    \begin{tabular}{lccccccccc}
    \toprule
	\multirow{2}*{\textbf{Method}} & \multicolumn{2}{c}{\textbf{VoxCeleb1-O}} & \multicolumn{2}{c}{\textbf{VoxSRC23 val}} \\
	\cmidrule(lr){2-3} \cmidrule(lr){4-5} \cmidrule(lr){6-7} & \textbf{EER[\%]} & \textbf{mDCF$_{0.01}$} & \textbf{EER[\%]} & \textbf{mDCF$_{0.05}$} \\
    \midrule 
     Base (Track1) & 0.622 & 0.058 & 3.727 & 0.207 \\
    + LMFT & 0.537 & 0.045 & 3.615 & 0.185 &  \\
    ++ AS-Norm & 0.489 & 0.047 & 3.037 & 0.159 & \\
    +++ QMF & 0.356 & 0.040 & 2.391 & 0.134 &   \\
    \midrule 
    Base (Track2) & 0.622 & 0.058 & 3.727 & 0.207 \\
    +Mix-FT (A+B) & 0.441 & 0.037 & 3.549 & 0.197  \\
    ++LMFT & 0.414 & 0.035 & 3.369 & 0.169 \\
    +++ AS-Norm & 0.356 & 0.037 & 3.041 & 0.148  \\
    ++++ QMF & 0.282 & 0.029 & 2.281 & 0.127  \\
    \bottomrule
    \end{tabular}
\end{table}

\subsection{Speaker Embedding Model}
Generally, the speaker embedding model consists of three parts: backbone network, encoding layer and loss function. The ArcFace \cite{arcface}, which could increase intra-speaker distances while ensuring inter-speaker compactness, is used as loss function. The backbone and encoding layer are reported as follows.

\subsubsection{Backbone Network}
In this module, we introduce three different speaker verification systems, including the ResNet100-based one \cite{idrd_voxsrc22}, the ResNet101-based one, 
and the ResNet152-based one. In addition, we also mount the SimAM \cite{simam} and fwSE \cite{idlab_voxsrc20} modules in backbone block. The acoustic features are 80-dimensional log Mel-filterbank energies with a frame length of 25ms and a hop size of 10ms. The extracted features are mean-normalized before feeding into the deep speaker network.

\subsubsection{Encoding Layer}
The encoding layer could transform the variable length feature maps into a fixed dimension vector, that is, the frame-level feature is converted to the segment-level feature. In this module, we adopt the temporal statistic pooling (TSP)\cite{xvector} and attentive statistic pooling (ASP) \cite{asp_pooling}.

\begin{table*}[htbp]\centering \scriptsize 
    \caption{The performance of various systems in the track1\&2. A and B in the column of Data ID denotes Vox2Dev and VoxBlink-clean, respectively.}
     \label{tab:final_system_task12}
    \begin{tabular}{llcccccccc}
    \toprule
    \multirow{2}*{\textbf{Task}} & \multirow{2}*{\textbf{ID \& Model}} & \multirow{2}*{\textbf{Channel}} & \textbf{Data} &  \multicolumn{2}{c}{\textbf{VoxCeleb1-O}} & \multicolumn{2}{c}{\textbf{VoxSRC23 val}} & \multicolumn{2}{c}{\textbf{VoxSRC23 test}} \\
    \cmidrule(lr){5-6} \cmidrule(lr){7-8} \cmidrule(lr){9-10} & & & \textbf{ID} & \textbf{EER[\%]} & \textbf{mDCF$_{0.01}$} & \textbf{EER[\%]} & \textbf{mDCF$_{0.05}$} & \textbf{EER[\%]} & \textbf{mDCF$_{0.05}$}  \\
    \midrule 
    \multirow{8}*{\textbf{Track 1}} & 1 SimAM-ResNet100-ASP & [64,128,256,512] & A & 0.356 & 0.040 & 2.391 & 0.134 & 2.285 & 0.132 \\
    & 2 SimAM-ResNet100-ASP(v2) & [128,128,256,256] & A & 0.404 & 0.029 & 2.519 & 0.141 & - & - \\
    & 3 ResNet100-TSP(v2) & [128,128,256,256] & A& 0.415 & 0.036 & 2.651 & 0.155 & - & - \\
    & 4 fwSE-ResNet100-ASP & [64,128,256,512] & A& 0.415 & 0.027 & 2.671 & 0.152 & - & - \\
    & 5 ResNet101-ASP & [64,128,256,512] & A& 0.388 & 0.021 & 2.395 & 0.128 & - & - \\
    & 6 ResNet152-ASP & [64,128,256,512] & A& 0.425 & 0.028 & 2.343 & 0.127 & - & - \\
    & 7 ResNet152-TSP & [64,128,256,512] & A& 0.394 & 0.030 & 2.551 & 0.153 & - & - \\
    \cmidrule(lr){2-10} & Fusion(1-7) & &   & 0.298 & 0.015 & 2.079 & 0.122 & 2.1910 & 0.1243 \\
    \midrule 
    \multirow{4}*{\textbf{Track 2}} & 8 SimAM-ResNet100-ASP & [64,128,256,512] & A+B & 0.282 &  0.029 & 2.281 & 0.127 & 2.350 & 0.126 \\
    & 9 SimAM-ResNet100-ASP(v2) & [128,128,256,256] & A+B & 0.319 & 0.047 & 2.395  & 0.145 & - - \\
    & 10 fwSE-ResNet100-ASP & [64,128,256,512] & A+B & 0.319  & 0.031 & 2.589  & 0.145 & - & - \\
    \cmidrule(lr){2-10} & Fusion(1-10) & &  & 0.218 & 0.013 & 2.029 & 0.114 & 2.0390 & 0.1165\\
    \bottomrule
    \end{tabular}
    \end{table*}

\subsection{Training Strategy and Implement Details}

We adopt the on-the-fly data augmentation \cite{on-the-fly} to add additive background noise or convolutional reverberation noise for the time-domain waveform. The MUSAN \cite{musan} and RIR Noise \cite{RIR} datasets are used as noise sources and room impulse response functions, respectively. To further diversify training samples, we apply amplification or playback speed change (pitch remains untouched) to audio signals. Also, we apply speaker augmentation with speed perturbation \cite{dku_voxsrc20}. We speed up or down each utterance by a factor of 0.9 or 1.1, yielding shifted pitch utterances that are considered from new speakers.

The speaker embedding model is trained through two stages. In the first stage, the baseline system adopts the SGD optimizer. We adopt the StepLR scheduler with 15 epochs decay. The init learning rate starts from 0.1, the minimum learning rate is 1.0e-4, and the decay factor is 0.1. The margin and scale of ArcFace are set as 0.2 and 32, respectively. We perform a linear warm-up learning rate schedule at the first 5 epochs to prevent model vibration and speed model training. The input frame length is fixed at 200 frames. Based on our previous study \cite{farfield_xiaoyi}, Mix-Finetune (Mix-FT) proves to be more effective than training from scratch by directly mixing data. Therefore, for Track 2, we inherite the models from Track 1 in the first stage and conducted Mix-FT by blending VoxBlink-clean data into the training set.  

In the second stage, we follow the SpeakIn training protocol\cite{speakin21} with large margin fine-tuning (LMFT) \cite{idlab_voxsrc20}. In the LMFT stage, we also remove the speaker augmentation. Moreover, the margin is increased from 0.2 to 0.4. According to the speaker embedding model size and the memory limit of GPU, the frame length is expended from 200 to 500. The learning rate decays from 5.0e-4 to 1.0e-5 under the 3 epochs.

\subsection{Score Calibration and Normalization}

The cosine similarity is adopted as the back-end scoring method. After scoring, results from all trials are subject to score normalization. We utilize Adaptive Symmetric Score Normalization (AS-Norm) \cite{asnorm} in our systems. The imposter cohort consists of the average of the length normalized utterance-based embeddings of each training speaker, i.e., a speaker-wise cohort with 5994 embeddings. The number of the top cohort is 300.

Quality Measure Functions (QMF) could calibrate the scores and improve the system's performance. For the QMF training set, we followed the trial structure rules of \cite{casv_xiaoyi} to construct Vox2dev based Vox-CA set, including Vox-CA5, Vox-CA10 and Vox-CA15, respectively. As described in \cite{idlab_voxsrc21,idrd_voxsrc22,dku_voxsrc22}, we adopt the following QMF qualities set to calibrate the scores processed by AS-Norm:
\begin{itemize}
    \item 1) logarithm the enrollment utterance's duration, 
    \item 2) logarithm the test utterance's duration,
    \item 3) magnitude of the enrollment embedding,
    \item 4) magnitude of the test embedding,
    \item 5) SNR of the enrollment utterance,
    \item 6) SNR of the test utterance,
    \item 7) verification score processed by the AS-Norm module. 
\end{itemize}
 
In this case, only SNR QMF 5) and 6) adopts the Max-Min normalization. The maximum and minimum values of SNE are obtained through the QMF training set. Logistic Regression is adopted to train the QMF. 

\subsection{Results}

\subsubsection{Baseline System}
In this subsection, we will show the performance of our baseline system and how we keep enhancing the performance step by step. We adopt the SimAM-ResNet100 as our baseline system for track 1. At first, we validate the baseline system under the VoxSRC23 validation and Vox1-O sets. Both LMFT and score calibration exhibit significant performance improvements, while QMF shows a relative gain of 20\%. For Track 2, we initially mix the VoxBlink-clean data with the training set, build upon the pre-trained model from Track 1, and then proceed with additional training stages. Subsequently, we perform the same LMFT and score calibration operations as the ones in Track 1.

\subsubsection{Fusion System}   
Table \ref{tab:final_system_task12} reports the final performance of all the systems we employed. All systems are fused at the scoring level. In Table 2, we observe that the performance on the validation set and the test set is nearly identical, indicating no overfitting. Therefore, we manually assigned weights to different models based on their performance on the validation set.

\section{Track 3: Semi-supervised domain adaptation}

Semi-supervised domain adaptation aims to adapt a well-trained source domain embedding extractor to the target domain by leveraging a limited amount of labeled data and a larger pool of unlabeled data from the target domain. 

\begin{figure*}
  \centering
  \includegraphics[width=\linewidth]{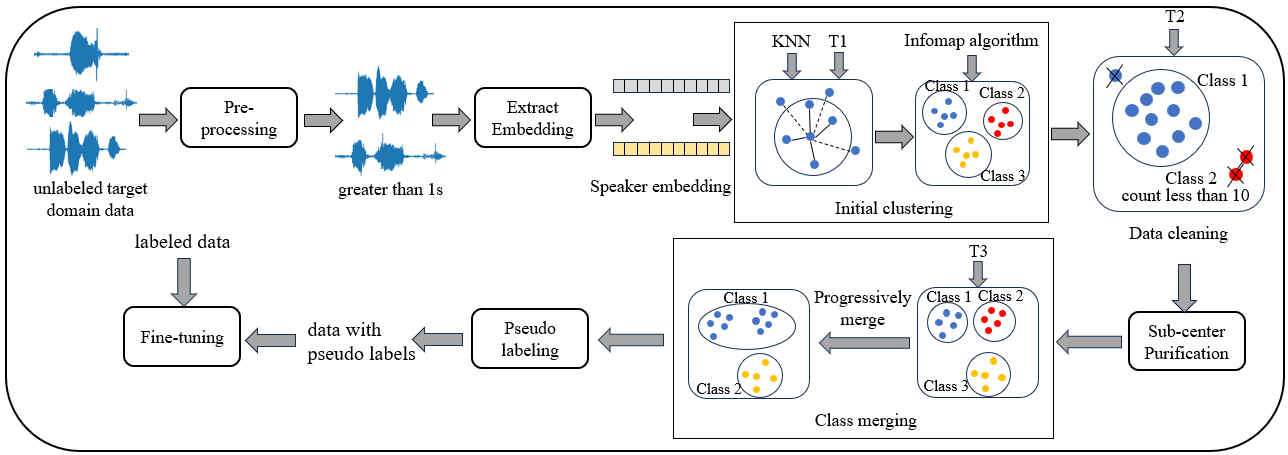}
  \caption{System framework of our pseudo-labeling method}
  \label{fig:system_framework}
\end{figure*}

\subsection{Pseudo Label based Method}
In this task, we propose a novel method for generating pseudo-labels in the context of semi-supervised learning. The overall approach involves first obtaining a series of highly pure classes, followed by the accurate merging of these classes to ultimately derive precise pseudo-labels. This method leverages the target domain information carried by the limited amount of labeled data to determine a set of thresholds for the initial clustering and class merging of unlabeled target domain data. Additionally, the Sub-center ArcFace \cite{subcenter-arcface} classifier is sensitive to noise, we utilize this characteristic to remove some noise classes. The system framework is show in Fig \ref{fig:system_framework}, and the following steps describe our pseudo-labeling based method:

\begin{itemize}
    \item Step 1. Pre-processing. Since audios with very short duration may not contain text information, only audios greater than 1s in length are retained.  
    \item Step 2. Extract embedding. Extract all speaker embeddings $\mathbf{Z} \in R^{N \times d}$ from the CN-Celeb dataset using the pre-trained speaker model.
    \item Step 3. Initial clustering. Generate the graph using the K-Nearest Neighbors (KNN) algorithm with a parameter $K$ set to 200. Remove edges with weights less than threshold $T1$. The Infomap \cite{infomap} algorithm is employed for initial clustering.
    \item Step 4. Data cleaning. Threshold $T2$ is adopted to remove outlier data within each class, and eliminate classes with a data count of less than 10.
    \item Step 5. Sub-center purification. Utilize the Sub-center ArcFace classifier's noise sensitivity to remove noisy classes.
    \item Step 6. Class merging. Progressively merge the classes using threshold $T3$.
    \item Step 7. Pseudo labeling. After completing the aforementioned steps, we ultimately obtained 1705 classes and 385356 utterances.
    \item Step 8. Fine-tuning. Both the unlabeled data with pseudo-labels and the labeled data are used for fine-tuning the speaker embedding model.
\end{itemize}

\subsection{Target Domain Threshold Determination}

Track 3 provides a limited amount of labeled target domain data containing 1,000 utterances from 50 speakers, so we want to extract some relevant domain information from it to improve pseudo-labeling for unlabeled target domain data. To achieve this, we first extract the speaker embeddings of the labeled target domain data using a well-trained source domain speaker embedding, and determine three thresholds from it. The purpose and the methods of setting these thresholds are explained as follows:

\textbf{Threshold} $\bm{T1}$ \textbf{.} Computing the cosine similarity between each embedding and all other embeddings, arranging these cosine similarity values in descending order, recording the cosine similarity value between each embedding and the first embedding with a different label, and selecting the maximum one as $T1$.

Purpose: When constructing a graph based on the KNN algorithm, it is possible to utilize $T1$ to remove some noisy edges.

\textbf{Threshold} $\bm{T2}$ \textbf{.} Calculating the cosine similarity between the embedding of each class and its respective centroid vector, then selecting the maximum value from the minimum similarity values of each class as $T2$.

Purpose: Removing outlier data for each class.

\textbf{Threshold} $\bm{T3}$ \textbf{.} Computing the cosine similarity between the centroid vectors of each class and selecting the maximum value as $T3$.

Purpose: As a limiting factor for merging classes.

\subsection{Sub-center Purification}
After obtaining the initial clustering results, examining the classes' purity before merging is essential. This step helps prevent erroneous merging due to the presence of outliers. In this process, we begin by assigning pseudo-labels to the data using the initial clustering results, using these as the input to train a Sub-center ArcFace classifier. After convergence, we stop training and pass all the data through the classifier to compute the selection probability for each class's sub-center. We consider that the most of the data from the highly pure class will tend to choose one specific sub-center, while the class with lower purity will exhibit multiple sub-centers and we could remove it.

\begin{table*}[htbp]\centering \scriptsize 
    \caption{The performance of various systems in the track3.}
     \label{tab:final_system_task3}
    \begin{tabular}{lccccccc}
    \toprule
    \multirow{2}*{\textbf{ID \& Model}} & \multicolumn{2}{c}{\textbf{VoxSRC23 val}} & \multicolumn{2}{c}{\textbf{VoxSRC23 test}} & \multicolumn{2}{c}{\textbf{VoxSRC22 test}}\\
    \cmidrule(lr){2-3} \cmidrule(lr){4-5} \cmidrule(lr){6-7}& \textbf{EER[\%]} & \textbf{mDCF$_{0.05}$} & \textbf{EER[\%]} & \textbf{mDCF$_{0.05}$} & \textbf{EER[\%]} & \textbf{mDCF$_{0.05}$} \\
    \midrule 
    1 SimAM-ResNet100-ASP & 7.490 & 0.342 & 5.287 & 0.3037 & 6.927 & 0.409 \\
    2 ResNet100-TSP(v2) & 7.350 & 0.360 & - & - & - & - \\
    3 SimAM-ResNet100-ASP(v2) & 7.525 & 0.335 & - & - & - & - \\
    4 ResNet152-ASP & 7.240 & 0.347 & - & - & - & - \\
    5 ResNet152-Stat & 7.535 & 0.358 & - & - & - & - \\
    \midrule
    Fusion(1+2) & 7.115 & 0.324 & 5.095 & 0.2869 & - & - \\
    Fusion(1+2+3+4+5) & 6.725 & 0.311 & 4.952 & 0.2777 & 6.584 & 0.374 \\
    \bottomrule
    \end{tabular}
\end{table*}

\subsection{Training Strategy and Implement Details}

In this task, the training consists of two parts: pre-training and fine-tuning. The pre-training stage is the same as the one in track1. We directly adopt the pre-trained models used in track1. In the fine-tuning stage, only target domain data applied speaker augmentation is used to fine-tune the pre-trained model. Since the distribution of the track3 validation set is the same as the evaluation set, QMF-based score calibration is trained on the validation set. For score normalization, 20,000 utterances are randomly selected as the cohort from the target domain unlabeled data. For long-duration utterances in the validation and evaluation set, we also truncate the audio for a maximum of 20 seconds.

\subsection{Results}

Table \ref{tab:final_system_task3} reports the system performance after score calibration in track3. The final submitted result achieves the EER of 4.952\% on the evaluation set. Additionally, we evaluate the performance on VoxSRC-22's track3 evaluation set, the result is 6.584\% EER, better than the first-place result last year (EER of 7.03\%).

\section{Conclusions}
This paper describes our systems in Track 1, 2 and 3. In Track 1, the cross-age QMF strategy aids in significantly improving the system performance of our Resnet-based model. For Track 2, we incorporate the VoxBlink-clean dataset and adopt a Mix-FT strategy for training. We also propose a novel method combining pseudo labeling with fine-grained label purification in Track 3. The final official submissions achieve an mDCF of 0.1243 for track 1, an mDCF of 0.1165 for track 2, and an EER of 4.952\% for track 3, respectively.

\bibliographystyle{IEEEtran}

\bibliography{mybib}

\end{document}